# Ant colonization processed algorithm for design of a toroidal shaped mobile 5G antenna

Sunit Shantanu Digamber Fulari

**Assistant professor at Govt college of Arts, Science and Commerce, Khandola, Marcela, Goa**
**Dept. of electronics and communication, Chandigarh University**

**Abstract:** There is great potential if we understand how nature functions, particularly the animals taking down from the ant to the larger animals. In this paper we will make an attempt to learn about ants colonization processing by studying their behaviour. Earlier there was particle swarm optimization which helped to solve many scientific problems. Ants communication with each other, their peculiar behaviour of working, colonization, their movements from one place to another, there is large potential in understanding their entire life to design a better 5G antenna.

**Index Terms:** Ant algorithms, Colonization, 5G mobile antenna.

**Introduction:** Ants colony establishment is a large topic with potential to solve many algorithmic problems in engineering and other fields too. Ant colonization is used to study the smart way of establishing a thing forward. Ants are known for storing food during the seasons when they can travel and utilize it during the times when it becomes tough to endure for food. This opportunity of finding food should be utilized in smart radiation when there is susceptibility of antennas operating with mobiles without wasting energy and reducing radiation.

Figure I: Ant movements in lines

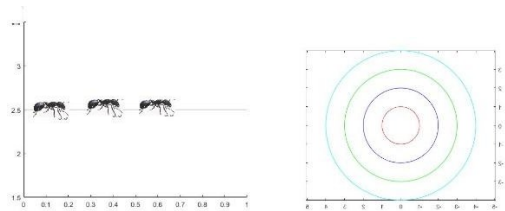

Ants travel by a very complex set of equations. They travel by their leader who leaves behind a trail of its journey. We will try to create an equation for this pattern of intermittent circles encircling into the center movement of ants as they enter into their homes.

Figure II: Smart ant path for radiation efficiency and toroidal antenna shape.

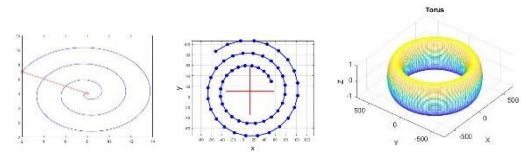

This is smart way of antenna radiation in which acoustic radiation is given out by the antenna in all three directions to determine how maximum efficiency can be attained. Antenna radiation should be monitored for maximum efficiency. Ants movement is what can be applied to our simulation.

Figure III: Acoustics of energy and shortest path.

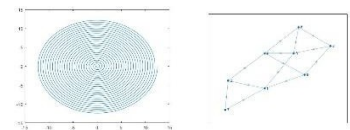

This pattern slowly covers the grasp and hold of the entire antenna by going in extended circles.

Antenna should function in such a way that it smartly covers acoustically a area separated by towers. Circular paths of radiation by antennas would be best understood by its nature of coverage of maximum area, as the figure below shows 3D coverage.

Figure IV: 3D radiation and two tower signal connection.

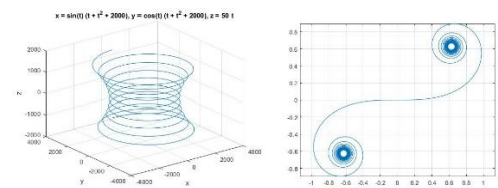

The figure below depicts how two colonies interact between each other, this ant colony algorithm should be utilized in the smart design of our antenna. The ants cover minimum area to travel from one colony to the other.

Torus shaped antenna is somewhat similar to our earlier design of circular antenna as it consists of intermittent connected circles.

Figure V:Signal excitation to Torus antenna.

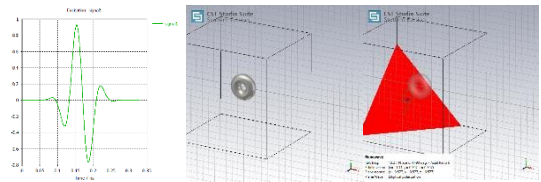

Torus is a shape which can easily socket into a cylindrical circular element. Torus with plane wave is used in our simulation. Waveguide port is also used.

figure VI: Far field pattern with waveguide.

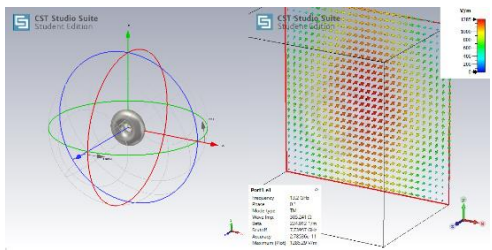

Figure VII:

S parameter plot for Torus antenna and impedance.

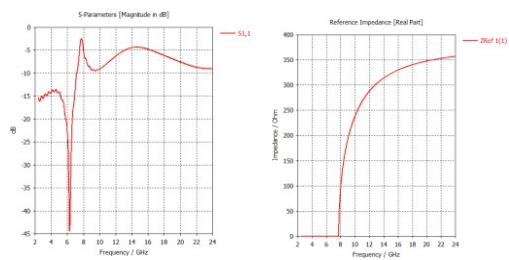

Figure VIII:Excitation power

Power watt varying from .5 to .05 with frequency.

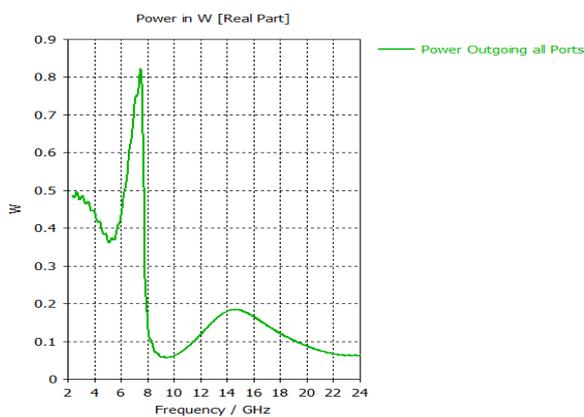

Figure IX and X:Field energy in magnitude and Voltage standing wave ratio.

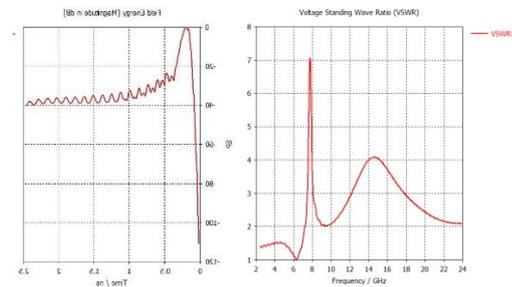

Figure XI and XII:Surface current for f=24GHz, zero phase and maximum(plot) 18.2186 A/m. Z parameter given by.

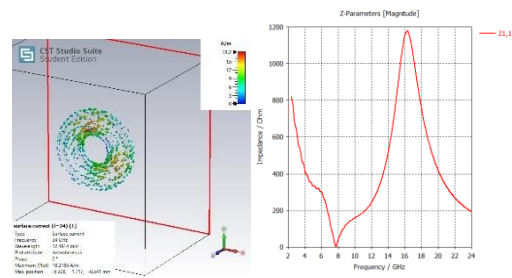

**II. Novelty of our work**:This torus design makes it very susceptible design in many ways to be comfortably fitted in easily with a end screw just like a bolt in driver. Torus design is itself of a bolt.

Our work hypothesis:Ant optimization using the colonization followed by ants is being studied to design a torus type antenna. Torus shaped antenna is very easy to configure, that is it can be easily rolled inside a cylindrical wire. The frequency chosen for operation of the antenna is from 2.4GHz to 24GHz.

Ants travel in straight line more or less very straight lines, colonization of ants and ants hill is one of theories of interest in our design of antenna. We want antennas to perform smartly just ascolonization of ants when they build a ant hill to store resources for the next season when there will be difficulties. Similarly the antennas require power to function, similarly it should also be able to store energy if required to work when there will be no power supply, in some areas where there is frequent power cuts and there is need for functioning of the antennas required.

Figure XII:Far Field pattern 4.5 and 9Ghz frequency is given in the figures as below

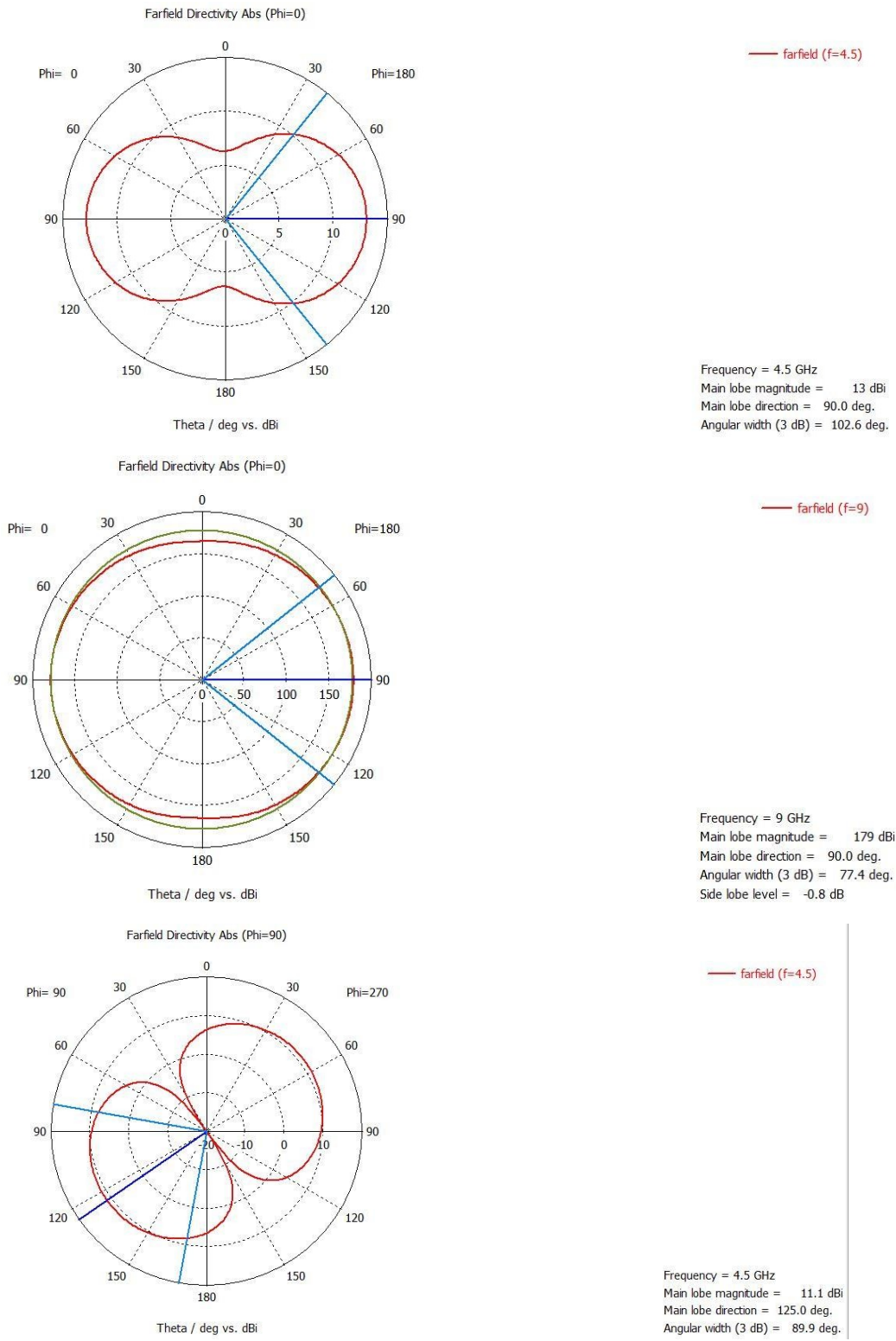

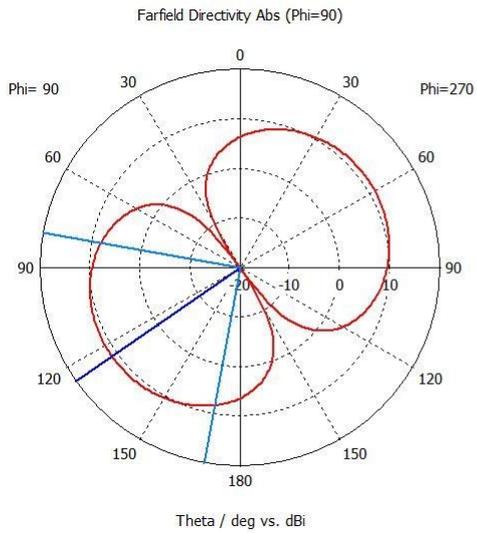

Frequency = 4.5 GHz
Main lobe magnitude = 11.1 dBi
Main lobe direction = 125.0 deg.
Angular width (3 dB) = 89.9 deg.

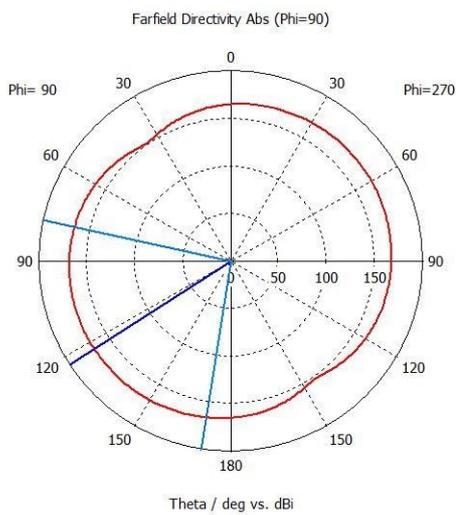

Frequency = 9 GHz
Main lobe magnitude = 169 dBi
Main lobe direction = 123.0 deg.
Angular width (3 dB) = 93.5 deg.

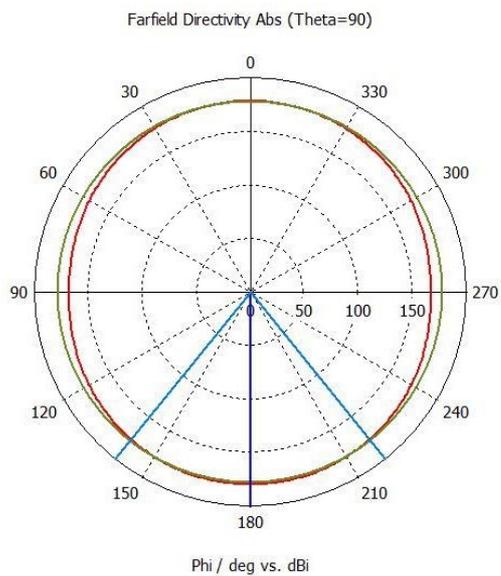

Frequency = 9 GHz
Main lobe magnitude = 179 dBi
Main lobe direction = 180.0 deg.
Angular width (3 dB) = 77.6 deg.
Side lobe level = -0.8 dB

**III. Benefits of this Torus shaped antenna:** The radiation pattern at 2.4,13.2 and 24GHZ shows that the antenna is good at radiating energy in all directions. Torus makes it behave like a bolt to be easily embedded into a cylindrical embedment. Primarily the antenna should be easily embedded and should perform the function of radiating energy. Energy dissipated in wrong direction is gone towaste and is also harmful for all forms of life.

**V. Our hypothesis based on Ant colonization:**

Ant colonization is a classic algorithm developed in 1990's which can be used to solve many problems. In this case the antenna is radiating energy very smartly,minimizing radiation levels, in addition to working with more power and reaching more area without consuming energy. The antenna will also function at higher frequencies varying from gigahertz from 2.4 to 24 in magnitude. What really this ant colonization contributes in solving our problem of smart radiation frequency is in efficiency, reaching more energy without more input and also functioning smartly.

**VI. Conclusion:** Toroidal shape is a very important subject of study as antenna and at the same time ant colonization algorithm is revisited. Ant colonization algotihm is used to solve the smart radiation problem in our letter. We have designed a torus antenna with back illuminated waveport. The frequency of operation was at high data rates varying from 2.4GHz to 24 GHz. The antennas was successfully shown good results for operation in mobile devices.

Declaration:

Funding:No funding received.

Conflict of interest:No conflict of interest.

Availability of data:No data procured.

Code availability:No code available.